\documentclass[12pt]{iopart}

\usepackage{iopams}

\usepackage{graphicx,epstopdf,epsfig}
\usepackage{color}
\usepackage{times}
\usepackage{hyperref}

\definecolor{violet}{rgb}{0.5,0.0,0.6}
\definecolor{blu}{rgb}{0.08,0.17,0.55}

\begin{document}
\title{Optical non-linearity in a dynamical Rydberg gas}
\author{J D Pritchard, A Gauguet, K J Weatherill and C S Adams}
\address{Department of Physics, Durham University, Rochester Building, South Road, Durham DH1~3LE, UK}
\ead{j.d.pritchard@durham.ac.uk}


\date{\today}

\begin{abstract}
We use the technique of electro-magnetically induced transparency (EIT) to probe the effect of attractive dipole-dipole interactions in a highly excited Rydberg gas.  The transient character of the EIT response is investigated by rapidly scanning the probe laser through resonance. We characterize the resulting cooperative optical non-linearity as a function of probe strength, density and scan direction. For the 58D$_{5/2}$ Rydberg state, an atom density of $1.6\times10^{10}~$cm$^{-3}$ and a positive frequency scan we measure a third-order non-linearity of $\chi^{(3)}=5\times 10^{-7}$~m$^2$V$^{-2}$. For the reverse scan we observe a second order non-linearity of  $\chi^{(2)}=5\times 10^{-6}$~mV$^{-1}$. The contrasting behaviour can be explained in terms of motional effects and resonant excitation of Rydberg pairs.
\end{abstract}

\submitto{\JPB}

\maketitle

\section{Introduction}
The excitation of laser cooled atoms to highly excited Rydberg states has opened many new research directions in quantum information \cite{saffman10}, many-body quantum physics \cite{honer10} and plasma physics \cite{killian07}. The use of electromagnetically induced transparency (EIT) to probe cold Rydberg gases \cite{weatherill08} provides a complementary technique to ionization for studying the properties of strongly interacting Rydberg gases \cite{raitzsch08} and opens new perspectives in non-linear optics \cite{pritchard10}. The effect of the strong dipole-dipole interactions between Rydberg atoms on EIT is to produce a non-local optical non-linearity \cite{ates11}. As such non-linearities can be large they are of considerable interest in the context of non-linear optics at the single photon level \cite{friedler05,shahmoon10}.

\begin{figure}
	\centering
	\includegraphics{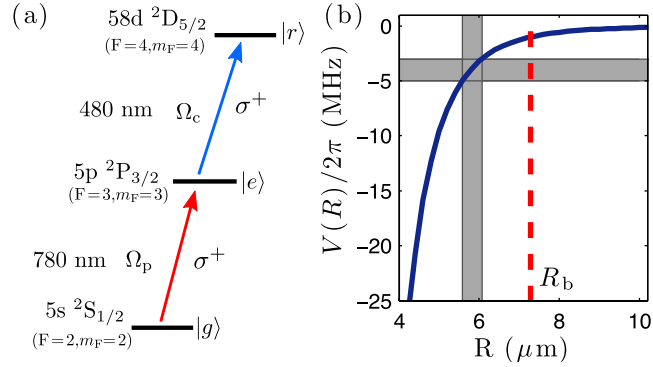}
	\caption{(a) Three-level atom system. The probe laser detuning $\Delta_\mathrm{p}=\omega_\mathrm{p}-\omega_{eg}$ is scanned across the $\vert g\rangle$ to $\vert e \rangle$ transition whilst keeping the coupling laser locked on resonance with $\vert e\rangle$ to $\vert r \rangle$. (b) Dipole-dipole interaction  potential for the 58$D_{5/2}$ pair state. The dashed line indicates the blockade radius $R_\mathrm{b}$ below which only a single atom can be excited to the Rydberg state on resonance. States with an energy shift of $-3$ to $-5$~MHz are highlighted, corresponding to an interatomic separation of $R=5.6-6.1~\mu$m.\label{fig:schematic}}	
\end{figure}

Electromagnetically induced transparency arises from interference between an atom and two coherent optical fields. A three-level atom with ground state $\vert g \rangle$, intermediate excited state $\vert e\rangle$ and Rydberg state $\vert r \rangle$, shown schematically in \fref{fig:schematic}~(a), is driven by a strong coupling laser and a weak probe laser with Rabi frequencies $\Omega_\mathrm{p}, \Omega_\mathrm{c}$ respectively. On the two-photon resonance, the atom evolves into a dark state,
\begin{equation}
\vert 0 \rangle = \frac{\Omega_\mathrm{c}\vert g \rangle - \Omega_\mathrm{p}\vert r \rangle}{\sqrt{\Omega_\mathrm{p}^2+\Omega_\mathrm{c}^2}},
\end{equation}
which no longer couples to the probe field, leading to transparency. For a pair of atoms separated by a distance $R$, dipole-dipole interactions between the Rydberg states cause a shift of the Rydberg pair-state energy level. For $V(R)>\gamma_\mathrm{EIT}$, where $\gamma_\mathrm{EIT}$ is the bandwidth of the EIT resonance, this leads to a blockade of excitation of the Rydberg pair state \cite{lukin01}. The blockade mechanism modifies the non-interacting two-atom dark state $\vert 0 \rangle$ as $\Omega_\mathrm{p}$ is increased, allowing population of $\vert{e}\rangle$ which resonant couples to the probe laser \cite{pritchard10,moller08}. The result is a suppression of transmission on the two-photon resonance, and hence a cooperative optical non-linearity mediated by interactions with surrounding atoms. This non-linearity is enhanced as more atoms are included in the blockade region which leads to a quadratic density dependence \cite{ates11}.

The optical response of Rydberg ensembles is sensitive to the nature of the dipole-dipole interactions between Rydberg atoms, which can be either repulsive or attractive and depend on the alignment angle between the dipoles. Rydberg interactions in cold gases has been the topic of in depth reviews, both experimental \cite{comparat10} and theoretical \cite{saffman10}. In previous work \cite{pritchard10} the focus was on $n$S$_{1/2}$ Rydberg states with zero orbital angular momentum, in the long range limit where the dipole dipole interaction is described by a repulsive, spherically symmetric van der Waals type interaction $V(R)=-C_6/R^6$. While this case allows a direct interpretation of the underlying physics (ruling out alternative mechanisms such as ionization), larger interactions and consequently larger optical non-linearities may be accessible by employing Rydberg states with non-zero angular momentum. In this paper we focus on the 58D$_{5/2}$ Rydberg state, coupling to the $M=m_{j1}+m_{j2}=5$ pair state which experiences a strong attractive interaction with a van der Waals interaction strength of $C_6\simeq150~\mathrm{GHz}~\mu$m$^6$ \cite{reinhard07} compared to $C_6\simeq -140~\mathrm{GHz}~\mu$m$^6$ \cite{singer05} for the nearby $60S_{1/2}$ state. A consequence of the attractive interactions is to introduce a temporal dependence into the EIT due to motional effects in the atomic cloud, which is explored through studying the spectra as a function of both frequency and time.

\section{Experiment}
To determine the optical non-linearity for the case of attractive interactions, we measure transmission through a laser cooled ensemble of $^{87}$Rb atoms using the experimental setup described in ref.~\cite{weatherill08}. Atoms are loaded into a magneto-optical trap (MOT), reaching a peak density of $\rho=1.6\times10^{10}$~cm$^{-3}$. A 10~ms optical molasses reduces the temperature to $20~\mu$K, after which the atoms are prepared in the 5s $^2$S$_{1/2}~(F=2,m_F=2)$ state ($\vert g \rangle$) by a 3~ms optical pumping pulse on the 5s $^2$S$_{1/2}$ $F=2$ $\rightarrow$ 5p $^2$P$_{3/2}$ $F'=2$ transition with $\sigma^+$ polarisation. The density of the atomic sample is controlled by turning off the repump light resonant with the 5s $^2$S$_{1/2}$ $F=1$ $\rightarrow$ 5p $^2$P$_{3/2}$ $F'=2$ transition during the optical pumping pulse, causing atoms to collect in the $F=1$ hyperfine ground state. This allows the density of atoms in $\vert g \rangle$ to be reduced without changing the cloud size. Following a 1~ms free expansion, EIT spectroscopy is performed using counter-propagating probe and coupling lasers, both $\sigma^+$-polarized. The coupling laser is a frequency doubled diode laser system at 480~nm, which is stabilized to the 5p $^2$P$_{3/2}~F'=3\rightarrow$ 58d~$^2$D$_{5/2}$ transition using EIT in a room temperature Rb cell \cite{mohapatra07,abel09}. This is focused to a $1/{\rm e}^2$ waist of $215\pm10~\mu$m with a power of 80~mW, giving a peak Rabi frequency of $\Omega_\mathrm{c}/2\pi=4.6$~MHz. The probe laser is scanned across the 5s $^2$S$_{1/2}~F=2$ ($\vert g \rangle$) $\rightarrow$ 5p $^2$P$_{3/2}~F'=3$ ($\vert e \rangle$) transition over a period of 960~$\mu$s. The detuning is linearly ramped either from $\Delta_\mathrm{p}/2\pi = -20\rightarrow+20\rightarrow-20$~MHz corresponding to negative-positive-negative (NPN) or $+20\rightarrow-20\rightarrow+20$~MHz corresponding to positive-negative-positive (PNP) frequency order. This double scan technique provides useful information on both atom loss \cite{weatherill08} and hysteresis in the experiment. The probe beam is focused to a waist of $160\pm10~\mu$m, with data recorded for powers varying from 5-40~nW corresponding to peak Rabi frequencies in the range $\Omega_\mathrm{p}/2\pi=0.3$--1~MHz. Transmission through the cloud is recorded using a fast photodiode with a 20~MHz bandwidth. The optical path length $\ell$ is determined from time of flight imaging of the cloud \cite{lett89}, giving $\ell = 0.9\pm0.1$~mm.

\section{Temporal dependence of the EIT lineshape}
Theoretical treatments of the optical non-linearity arising from dipole-dipole interactions in EIT \cite{ates11,sevilay11} have so far assumed the sample to be frozen, with no modification of the interatomic separations in the atom cloud due to interactions.  However, even at 20~$\mu$K the motional effects from the attractive interactions can be significant on timescales of tens of microseconds \cite{amthor07,amthor09}. Figure~\ref{fig:schematic}~(b) shows the dipole-dipole interaction potential $V(R)$ for the 58$D_{5/2}$ pair state as a function of interatomic separation. For the experiments presented below, $\gamma_\mathrm{EIT}/2\pi\sim1$~MHz corresponding to a blockade radius of $R_\mathrm{b}=\sqrt[6]{C_6/\gamma_\mathrm{EIT}}\simeq7~\mu$m which is indicated by a dashed line. Comparing this to the average interatomic separation given by $\langle R \rangle = (5/9)\rho^{-1/3}$ \cite{hertz09}, for a density of $10^{10}$~cm$^{-3}$ $\langle R \rangle\sim2.5~\mu$m so the blockade effect is expected to be clearly observable for the experiment parameters. For atoms initially separated by $7~\mu$m, the collision time is approximately 20~$\mu$s \cite{amthor09}, resulting in ionisation of at least one of the two atoms. These two effects of motion and ionisation introduce a temporal dependence into the EIT spectra, making the optical  non-linearity a function of not only optical fields, but also the timescale over which they are probed. 

In addition to the resonant blockade behaviour, the EIT spectra are also strongly influenced by the off-resonant excitation of the anti-blockaded pair states at $\Delta_\mathrm{p}=V(R)/2$ that have a separation $R<R_\mathrm{b}$, and therefore ionise more rapidly. As an example, states with $V(R)/2\pi=-3$ to $-5$~MHz (corresponding to a detuning of $\Delta_\mathrm{p}/2\pi=-1.5$ to $-2.5$~MHz) are highlighted in figure~\ref{fig:schematic}~(b), with interatomic separations of $R\sim5.8~\mu$m, with a collision time of around 8~$\mu$s. This has two effects, firstly creating ions off-resonance that can then affect the behaviour on the EIT resonance, and secondly it reduces the number of short-range pairs in the medium. Off-resonant excitation of these anti-blockade states has previously been used as a method of mapping out the distribution of nearest-neighbours within the atom cloud \cite{amthor10}.

\begin{figure}[t!]
	\centering
	\includegraphics{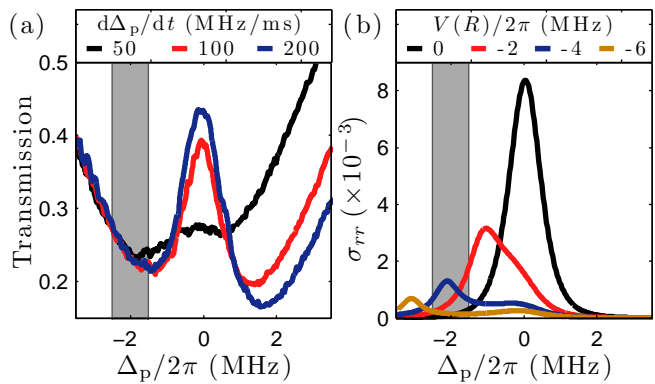} 
	\caption{Temporal effects in EIT. (a) Measured EIT spectra recorded at 30~nW ($\Omega_\mathrm{p}/2\pi=0.9$~MHz) for a range of scan speeds. For the slowest scan (50~MHz/ms) there is evidence of broadening and loss consistent with ionization or motional effects starting at the detuning corresponding to pair states highlighted in figure~\ref{fig:schematic}~(b), whilst at faster speeds the spectra appear consistent. Fitting the high-scan rate data gives $\Omega_\mathrm{c}/2\pi=2.4$~MHz and a relative two-photon linewidth of 200~kHz. (b) Rydberg pair population calculated using the pair-model described in ref.~\cite{pritchard10} calculated for different interaction strengths. This shows resonant excitation of the pairs with $R<R_\mathrm{b}$ before scanning across resonance, explaining the ionisation in (a). \label{fig:temporal}} 
\end{figure}

To demonstrate these temporal effects, EIT spectra are recorded at a density of $\rho=0.9\times10^{10}$~cm$^{-3}$ with $\Omega_\mathrm{p}/2\pi=0.9~$MHz for a positive scan across the resonance for a range of scan speeds, shown in figure~\ref{fig:temporal}~(a). For the slowest frequency ramp of 50~MHz/ms, the laser scans across the EIT resonance in 25~$\mu$s. Around $\Delta_\mathrm{p}/2\pi=-2$~MHz, corresponding to the excitation of the anti-blockaded states highlighted in figure~\ref{fig:schematic}~(b), the transmission increases giving a broadened and very asymmetric EIT resonance which has a narrow absorption profile, with atoms appearing to be lost rapidly around $\Delta_\mathrm{p}/2\pi\sim1$~MHz. This spectrum is consistent with significant ionisation of the off-resonant pair states, which appears to dominate over the EIT suppression expected due to blockade, seen from a broadening of the EIT profile. Figure~\ref{fig:temporal}~(b) shows calculations of the pair-state population using the pair model detailed in ref.~\cite{pritchard10} as a function of interaction strength for $\Omega_\mathrm{c}/2\pi=2.4$~MHz to match the experiment parameters. This demonstrates the suppression of the resonant pair excitation due to blockade, whilst highlighting importance of the off-resonant excitation of the anti-blockade states for negative detunings. For large interactions, the off-resonant pair excitation is suppressed due to being far-detuned, which is why the effects of the pair states only become observable around $\Delta_\mathrm{p}/2\pi\sim-2$~MHz despite the average interatomic separation giving a shift of several hundred MHz. From the groundstate density, the probability of finding atoms separated by 5.8~$\mu$m is around 0.1~\%, corresponding to a pair state density around $10^{4}$~cm$^{-3}$ at this detuning, which can then ionise and broaden the resonance. As the scan speed is increased, the spectra in figure~\ref{fig:temporal}~(a) become symmetric with approximately constant transmission on resonance, showing that loss due to ionization is suppressed as there is not enough time for a significant number of ions to be created during the scan across resonance. All data presented below is recorded for a scan speed of 80~MHz/ms, fast enough that there is no evidence of loss or asymmetry across the first EIT resonance but slow enough to enable these dynamic effects to be relevant. 

Despite there being no observable loss or asymmetry across the first resonance at this intermediate scan speed, there may be a small ion fraction in the cloud. For S-states as studied in \cite{pritchard10}, it is easy to assess the role of ionization in the suppression data as this leads to a red-shift of the EIT resonance. For the $D$-states however, the presence of an ion in the sample breaks the quantisation along the $z$-axis and will project the Rydberg atom into one of the $\vert m_j\vert=1/2$, 3/2 or 5/2 states which have scalar polarizabilities of $\alpha_0=-137$, 111, 607 MHz/(V/cm)$^2$ respectively, calculated using the Numerov method \cite{zimmerman79}. To estimate the effect this mixture of polarizabilities has on the lineshape, a model was developed to calculate the lineshape for a uniform density of atoms, from which a fixed fraction were chosen as ions. For each of the remaining atoms, the total field electric field due to the surrounding ions is used to find the Stark-shift assuming a random choice of $\vert m_j \vert$, and the total lineshape calculated by summing over the susceptibility of each atom. 
\begin{figure}[t!]
	\centering
	\includegraphics{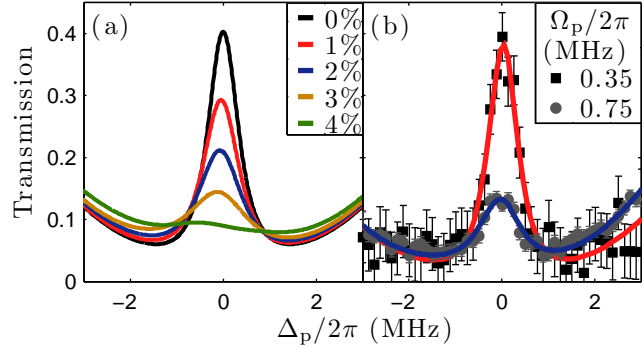}
	\caption{Role of ions in EIT suppression. (a) Simulated lineshape as a function of ion fraction for a random atomic distribution. Due to the sign difference in $\alpha_0$ of the different $m_j$ states, the lineshape isn't simply red-shifted as expected for S$_{1/2}$ states but instead the resonance disappears, becoming asymmetric in the wings. (b) Transmission data recorded at peak density showing strong suppression of the EIT for a strong probe as expected from interactions, however this shows good agreement with the ion model for a 3~\% yield.\label{fig:ion}}
\end{figure}

The results of the ion model are plotted in \fref{fig:ion} (a) as a function of ion density within the cloud. This shows that even for a very weak ion yield, the Stark-shift of the two-photon resonance significantly modifies the resonant transmission. This doesn't simply correspond to a red-shift, instead the line is split due to the varying $m_j$ components, reducing the transmission whilst becoming increasingly asymmetry either side of the original resonance. This is contrasted in \fref{fig:ion} (b) by experiments performed for a positive frequency scan at the peak density of $\rho=1.6\times10^{10}$~cm$^{-3}$ where ionization effects should be maximized. For the weak probe data the EIT lineshape is unperturbed, whilst at $\Omega_\mathrm{p}/2\pi = 0.7$~MHz the transmission feature is strongly suppressed as expected due to interactions. However, it looks similar to the Stark detuned profile expected for a 3~\% ion fraction. One caveat to this comparison is that the model assumes ions are present at all times in the scan, and not created dynamically. As seen from figure~\ref{fig:temporal}~(b) the off-resonant pair-states, and hence ions, are most likely to be excited close to the EIT resonance. In addition, the blockade mechanism suppresses the excitation of close Rydberg pairs which should prevent ion generation during the scan. As there is no way of measuring the ion yield in the current experiment, it is difficult to assess the relative importance of ionization over blockade for the positive scan data, however it is still interesting to consider a system that gives a non-linear optical response due to other mechanisms such as ionization. This could be extended to the idea of performing EIT on highly excited Rydberg states and using the Stark-shift to create an ion-blockade \cite{chotia08} which would switch the system from transmission to absorption, making a sensitive ion detector.

\section{Frequency dependence of EIT suppression}

\begin{figure}[b!] 
	\centering
	\includegraphics{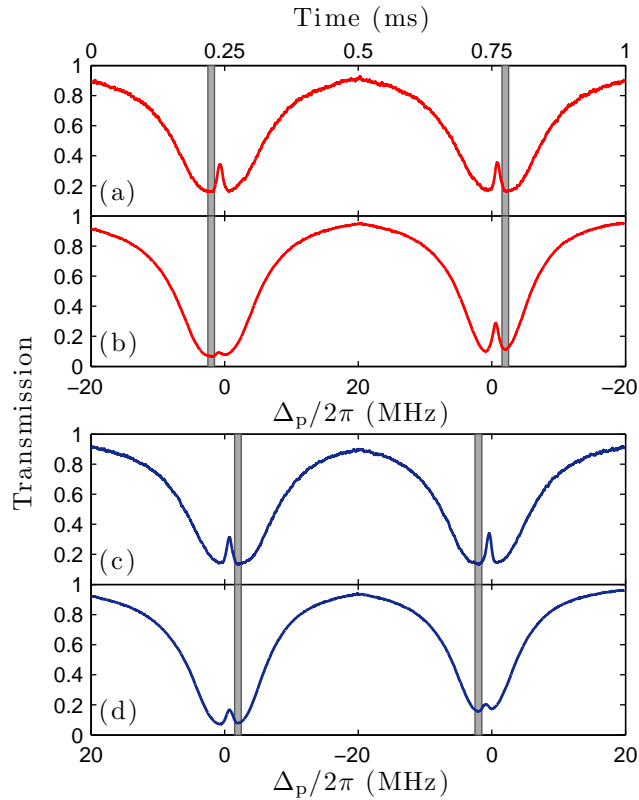}
	\caption{Rydberg EIT spectra for the $5S_{1/2}\rightarrow5P_{3/2}\rightarrow58D_{5/2}$ system for NPN (a) and (b), and PNP (c) and (d) scan directions at probe fields of $\Omega_\mathrm{p}/2\pi=0.3$~MHz (a) and (c) and 0.9~MHz (b) and (d). For the weak probe data (a) and (c) symmetric EIT is observed on the two photon resonances. At higher probe Rabi frequency (b) and (d), there is evidence of suppression of the EIT peak and asymmetry dependent upon the direction of the scan through resonance. For a positive scan there is a strong suppression, which is revived in the reverse scan at later time (b). The negative scan in (d) reveals loss in the second scan. For all traces, the frequency corresponding to excitation of the anti-blockade states from figure~\ref{fig:schematic}~(b) is highlighted. \label{fig:NPNvsPNP}}
\end{figure}

To analyse the dynamical response further we compare the behaviour of a positive frequency scan to the negative frequency scan. \Fref{fig:NPNvsPNP} shows data recorded for both  NPN (a) and (b) and PNP (c) and (d) scan directions for weak $\Omega_\mathrm{p}/2\pi=0.3$~MHz (a) and (c) and strong 0.9~MHz probe fields (b) and (d) at the peak density of $\rho=1.6\times10^{10}$~cm$^{-3}$. In the weak-probe regime the dark-state $\vert 0 \rangle\equiv\vert g \rangle$ and the interaction effects play no role in the behaviour of the system. Comparing data in (a) and (c), the scans show symmetric EIT traces with no hysteresis between the first and second scan, as expected. This confirms the ground-state density is not varying over the scan duration. For the strong probe data for the NPN scan direction in (b), the first resonance is strongly suppressed as observed above, however the reverse scan in the negative direction shows almost complete recovery of the EIT comparable to the weak-probe trace in (a), with a slight reduction in the optical depth relative to the first scan. Contrasting this to the opposite PNP case (d), on the initial scan the transmission is suppressed but only by around half that observed in the NPN data. The second scan in the positive frequency direction however shows significantly greater suppression of the resonance, in addition to around a 10~\% change in the background absorption from loss of atoms in $\vert g \rangle$. As well as the line-shape, the frequency of the two-photon resonance was compared for 20 datasets for a variety of probe powers. This shows no systematic shift for the strong probe data over the $\pm300$~kHz noise observed in the weak probe limit arising from fluctuations in the frequency of the coupling laser.

This data reveals the anti-blockade states play an important role in determining the magnitude of the EIT suppression, with the short-range pairs identified in figure~\ref{fig:schematic}~(b) highlighted here in grey for each scan. This clearly shows that the EIT suppression for the strong probe data of (b) and (d) is significantly larger when scanning from negative detuning across the short-range pair states compared to the suppression scanning from positive detuning. The enhancement is most likely due to ionisation of the short range pairs causing an irreversible ion-blockade to dominate over the coherent excitation blockade due to the longer range of the Coulomb potential. The pair-state ionisation also explains the hysteresis in the datasets, as for (b) the NPN scan first creates a strong suppression by excitation and ionisation of the short-range pair states. This reduces the number of atoms with separations of $R<R_\mathrm{b}$, leading to the observed recovery in the reverse scan 0.5~ms later. This hysteresis in the strong probe data is interesting, as the optical response is now dependent upon the experiment sequence modifying the atomic distribution. This is similar to direct observations of ion-yield for attractive states excited with either a red or blue detuning relative to the Rydberg energy \cite{amthor07,li06}.

\section{Optical Non-linearity}

As seen from the data in figure~\ref{fig:NPNvsPNP} the suppression of the first scan for the PNP data is due to the dipole-blockade mechanism, whilst for the NPN scan it appears to be enhanced by ionisation of the short-range pair states. To characterise the signatures associated with these two regimes, the magnitude and density dependence of the resonant optical non-linearity is measured. Data is taken for a range of probe powers and densities for both scan directions. For each set of parameters, a total of 20 traces are used to determine the transmission $T$ on the first EIT resonance. This is converted to the complex part of the susceptibility at the probe frequency using the relation $\mathrm{Im}\{\chi\}=-\log_e(T)/k\ell$, where $k=2\pi/\lambda_\mathrm{p}$ is the wavevector of the probe laser. 

\begin{figure}[t]
	\centering
	\includegraphics{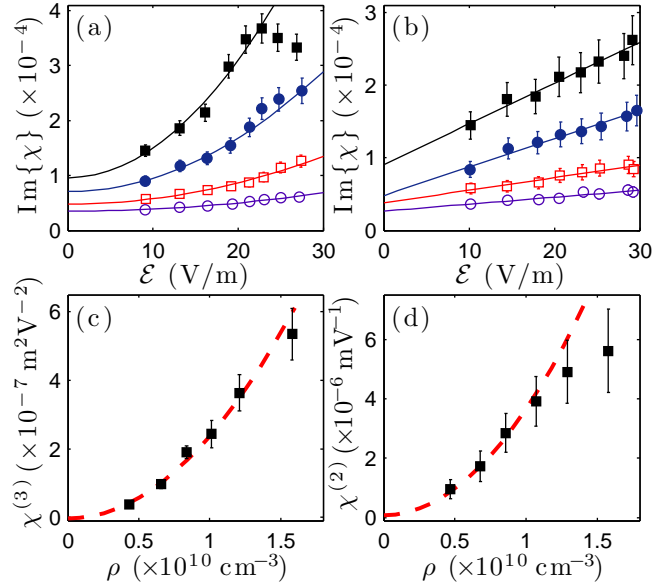}
	\caption{Non-linear susceptibility. Imaginary part of the susceptibility as a function of the probe electric field $\mathcal{E}$ for the positive (a) and negative (b) scan directions for $\rho$ = 0.4 ({\color{violet}{$\opencircle\!$}}) 0.7 ({\color{red}{$\opensquare\!$}}) 1.0 ({\color{blu}{$\fullcircle\!$}}) 1.6 ($\fullsquare$) $\times10^{10}$~cm$^{-3}$. Data in (a) and (b) is fitted to third- and second-order non-linearities respectively, and the resulting density dependence plotted in (c) and (d). Both $\chi^{(2)}$ and $\chi^{(3)}$ display a quadratic density scaling, consistent with pair-wise interactions.\label{fig:chi}}	
\end{figure}

The measured susceptibility is plotted as a function of the probe electric field $\mathcal{E}$ in \fref{fig:chi}, for both the NPN (a) and PNP (b) scan directions. In the high-density data in (a) the non-linearity is seen to saturate around 20~V/m ($\Omega_\mathrm{p}/2\pi=0.8$~MHz), after which it reverses direction. This turning point appears due to loss of atoms as the laser scans across the resonance, and places a limit on the largest probe power that can be used. All of the curves in (a) scale consistent with a third-order non-linearity. The function $\chi\!=\!\chi^{(1)}\!+\!\chi^{(3)}\mathcal{E}^2$ is fit to all the datasets, giving a peak value of $\sim5\times 10^{-7}$~m$^2$V$^{-2}$ for the highest density data. This represents a very large non-linearity for an atomic sample, comparable in magnitude to that observed in slow light experiments at BEC densities \cite{hau99}. For the PNP data in (b), the maximum suppression of the resonant transmission is lower, and there is no evidence of saturation even at strong fields and high density. The susceptibility reveals a linear scaling with $\mathcal{E}$ consistent with a second order response and is fit to $\chi\!=\!\chi^{(1)}\!+\!\chi^{(2)}\mathcal{E}$, giving a peak value of $\chi^{(2)}=5\times 10^{-6}$~mV$^{-1}$. This difference in the non-linear scaling for the two scan directions is consistent with there being different mechanisms (blockade or ionisation) responsible for the suppression for each scan direction, as discussed above.

Comparing this data to the theoretical results obtained from the many-body Monte-Carlo model of Ates {\it et al.} \cite{ates11}, the model predicts a third-order non-linearity which saturates as $\Omega_\mathrm{p}$ approaches $\Omega_\mathrm{c}$ for the suppression due to blockade from repulsive interactions. However, from the data a third order non-linearity is observed for the case of enhancement due to ionisation, whilst the suppression due to dipole blockade gives a second order scaling. This discrepancy requires a more detailed study involving direct measurement of the ion fraction for each scan direction to explore the importance of these temporal effects compared to the theory that considers a static atomic cloud to try and elucidate the reason for this difference in the order of the optical non-linearity. Despite the data in (a) and (b) showing different non-linear scalings, the maximum values of the complex part of the susceptibility for a given density are comparable. This is limited by two factors - the first is the absolute optical depth of the sample in the absence of the coupling laser, as this sets the maximum value of $\chi$ in the experiment, and the second is the number of atoms per blockade sphere, $\mathcal{N}$. As more atoms are added to the blockade, the dark-state fraction reduces proportional to 1/$\mathcal{N}$, leading to a non-linear density dependence. The density dependence of the non-linear susceptiblity is shown in (c) and (d) for the NPN and PNP data respectively. This reveals a clear quadratic scaling of $\chi^{(3)}$ with density and an initial quadratic dependence for $\chi^{(2)}$ that saturates at higher densities. These observations are consistent with the expected density scaling found in ref.~\cite{ates11}.

\section{Conclusions and Outlook}
In summary, we have shown that the optical response observed for EIT in an attractive gas is strongly dependent upon the hysteresis in the system due to motional effects and ionisation of the red-detuned anti-blockaded states. For a positive frequency sweep, the resulting third-order non-linearity is very large due to the enhancement from the ionisation of the short-range pair states. A negative frequency sweep however reveals a second-order non-linearity due to the blockade effect.

In future work, the attractive potential will be characterized in the dispersive regime for a single blockade volume to explore its application to single-photon quantum gates. An additional parameter accessible for non-zero angular momentum states is the dependence on the alignment of the dipoles \cite{carrol04}. This could be explored with the application of a weak electric field, potentially enabling an external switch to turn the blockade-mechanism on or off.
\ack
We are grateful to I G Hughes and M P A Jones for stimulating discussions and thank Durham University and EPSRC for financial support.

\end{document}